\documentclass[namedreferences]{kluwer}

\usepackage[dvips]{graphicx}

\begin{document}
\begin{article}

\begin{opening}

	\title{Influence of interstellar and atmospheric extinction on light curves of eclipsing
	       binaries}
	\author{A. \surname{Pr\v sa}}
	\author{T. \surname{Zwitter}}
	\institute{University of Ljubljana, Dept. of Physics, Jadranska 19, 1000
	           Ljubljana, Slovenia}

\begin{ao}
	andrej.prsa@fmf.uni-lj.si
\end{ao}

\begin{abstract} 
Interstellar and atmospheric extinctions redden the observational photometric data and they should be handled rigorously. This paper simulates the effect of reddening for the modest case of two main sequence $T_1 = 6500$K and $T_2 = 5500$K components of a detached eclipsing binary system. It is shown that simply subtracting a constant from measured magnitudes (the approach often used in the field of eclipsing binaries) to account for reddening should be avoided. Simplified treatment of the reddening introduces systematics that reaches $\sim 0.01$mag for the simulated case, but can be as high as $\sim 0.2$mag for e.g. B8~V--K4~III systems. With rigorous treatment, it is possible to \emph{uniquely} determine the color excess value $E (\mathrm B - \mathrm V)$ from multi-color photometric light curves of eclipsing binaries.
\end{abstract}


\end{opening}

\section{Introduction}

Although interstellar extinction has been discussed in many papers and quantitatively determined by dedicated missions (IUE, 2MASS and others), there is a lack of proper handling in the field of eclipsing binaries. The usually adopted approach is to calculate the amount of reddening from the observed object's coordinates and its inferred distance and to subtract it uniformly, regardless of phase, from photometric observations. This paper shows why this approach may be inadequate, especially for objects where interstellar extinction and the color difference between both components are significant. Atmospheric extinction is a better-posed problem: similarly as interstellar extinction depends on $E (\mathrm B - \mathrm V)$, atmospheric extinction depends on air-mass, which is a measurable quantity, whereas $E (\mathrm B - \mathrm V)$ has to be estimated.

\section{Simulation}

To estimate the effect of reddening on eclipsing binaries, we built a synthetic binary star model, consisting of two main sequence G9~V--F5~V stars with $T_1 = 5500K$, $R_1 = 0.861 R_\odot$ and $T_2 = 6500K$, $R_2 = 1.356 R_\odot$ and 1 day orbital period. The simulation logic is as follows: for the given phase, we calculate the effective spectrum of the binary by convolving Doppler-shifted individual spectra of the visible surfaces of both components. To this intrinsic spectrum we rigorously apply interstellar and atmospheric extinctions (both as functions of wavelength). We then convolve this reddened spectrum with instrumental response function (composed of the filter transmittivity and detector response functions) and integrate over the bandpass wavelength range to obtain the flux. In contrast, we use the same intrinsic spectrum without rigorously applying the reddening. To simulate the subtraction\footnote{To emphasize that this constant is \emph{subtracted} from photometric observations (given in magnitudes), we shall use the term \emph{constant subtraction} throughout this study, although we are \emph{dividing} when operating with fluxes.} of a reddening \emph{constant} from photometric observations, we simply divide the intrinsic spectrum by the flux that corresponds to this constant. Finally, we calculate the flux in the same manner as before and compare it to the flux obtained by applying rigorous reddening.

For building synthetic light curves we use {\tt PHOEBE}\footnote{{\tt PHOEBE} stands for PHysics Of Eclipsing BinariEs and it is based on the Wilson-Devinney code. See {\tt http://www.fiz.uni-lj.si/phoebe} for details.} (Pr\v sa \& Zwitter, 2004; in preparation). Each light curve consists of 300 points uniformly distributed over the whole orbital phase range. To be able to evaluate the impact of reddening on photometric light curves exclusively, all second-order effects (limb darkening, gravity brightening, reflection effect) have been turned off.

\begin{figure}
\begin{center}
\includegraphics[width=10cm,height=5cm]{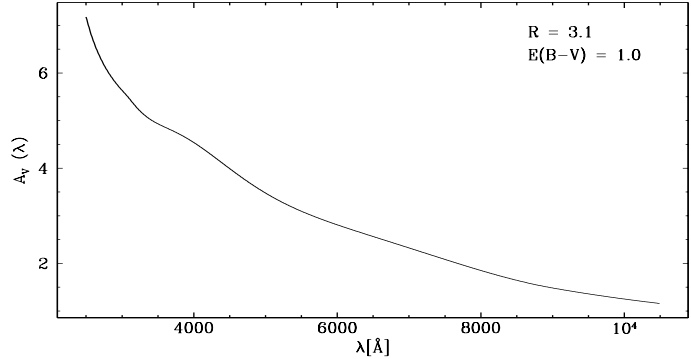}
\caption{The reddening law adopted from Cardelli et al.~(1989).}
\label{redlaw}
\end{center}
\end{figure}

\begin{figure}
\includegraphics[width=6cm,height=3cm]{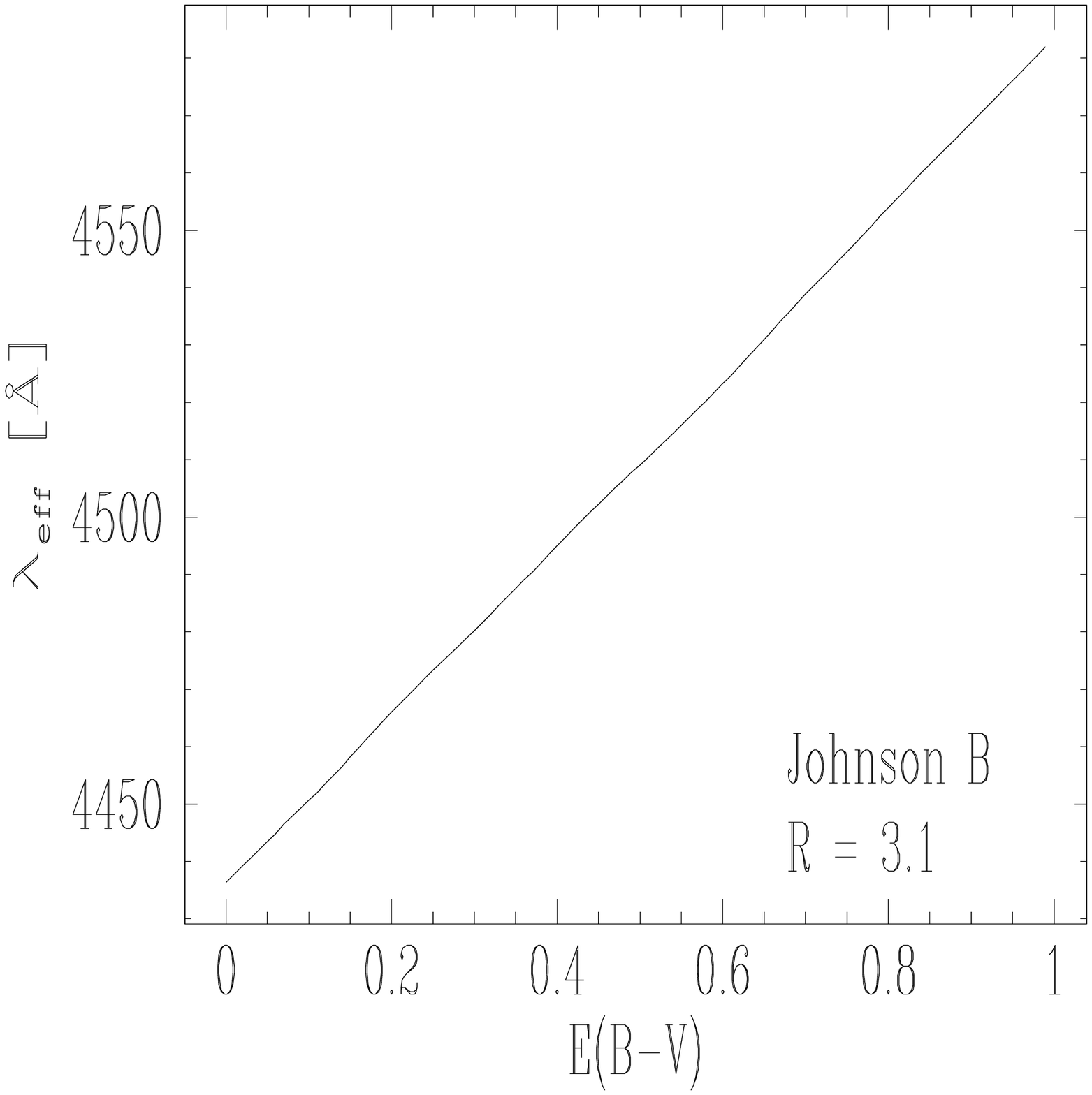}
\includegraphics[width=6cm,height=3cm]{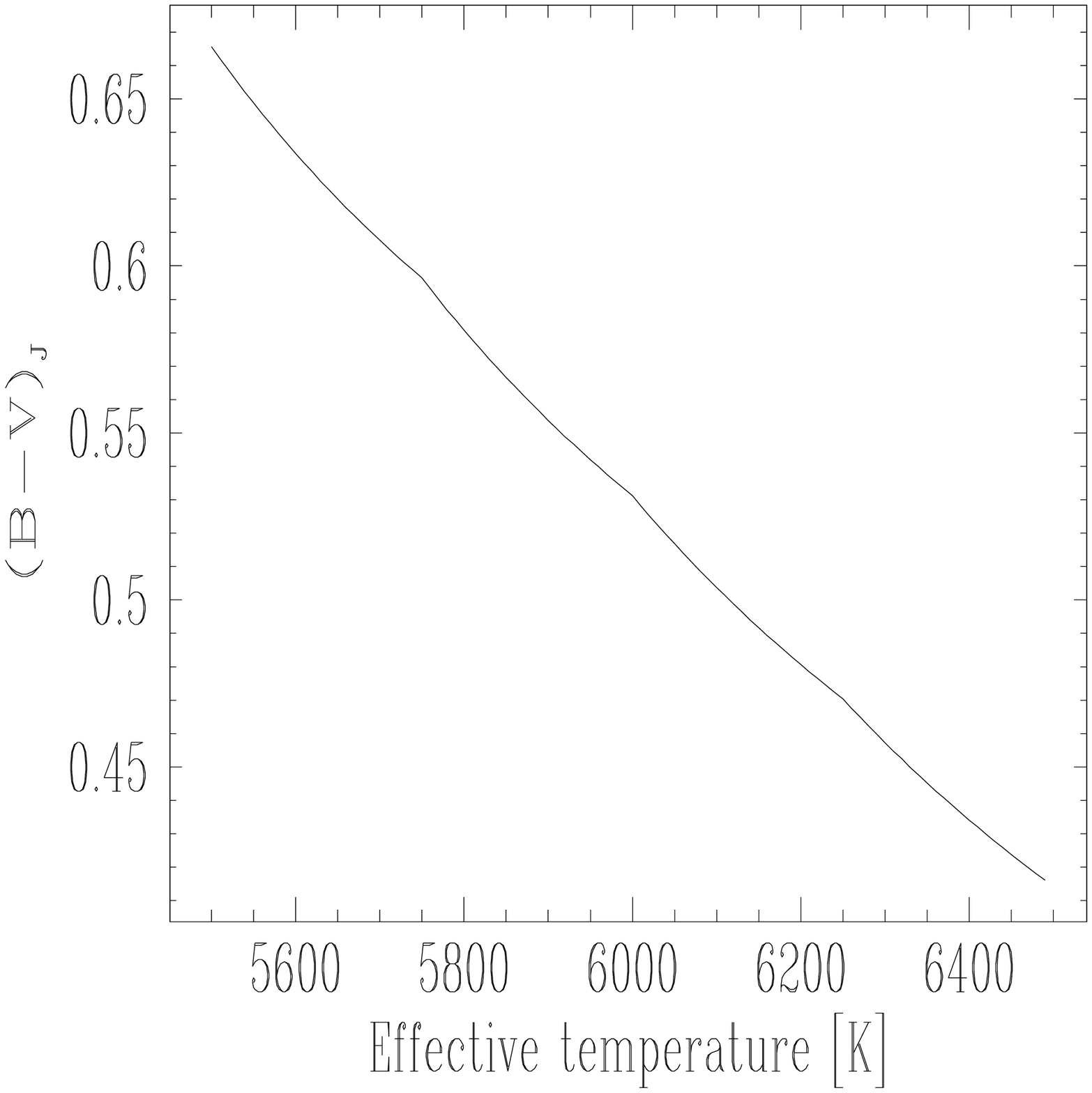} \\
\caption{Left: The change of the effective wavelength of the Johnson B filter due to reddening of the simulated G9~V--F5~V binary. Right: The $(\mathrm B-\mathrm V)_{\mathrm J}$ color index on the G9~V--F5~V (5500~K--6500~K) temperature interval, calculated by integrating the spectrum over both filter bandpasses.}
\label{weff}
\end{figure}

We take Kurucz's synthetic spectra ($R=20000$) from precalculated tables by Munari et al.~(2004; in preparation). The used $(UBV)_\mathrm{J} (RI)_\mathrm{C}$ response data (filter $\times$ detector) are taken from ADPS \cite{moro2000}, where we apply a cubic spline fit to obtain the instrumental response function.

For interstellar extinction, we use the \inlinecite{cardelli1989} empirical formula (Fig.~\ref{redlaw}), where $\mathcal R_\mathrm V = 3.1$ was assumed throughout this study. \inlinecite{schlegel1998} interstellar dust catalog was used to obtain the maximum color excess $E(\mathrm B - \mathrm V)$ values for different lines of sight.

For atmospheric extinction we use the equation triplet for Rayleigh-ozone-aerosol extrinction sources given by \inlinecite{forbes1996} and summarized by \inlinecite{pakstiene2003}. The observatory altitude $h=0$~km and the zenith air-mass are assumed throughout the study.

To rigorously deredden the observations for the given $\mathcal R_V$ and $E(\mathrm B - \mathrm V)$, it is necessary to determine the reddening for each wavelength of the spectrum. Correcting differentially and integrating over the filter bandpass then yields the dereddened flux of the given filter. However, without spectral observations, it is difficult to calculate properly the flux correction. Since \inlinecite{cardelli1989} formula depends on the wavelength, the usually adopted approach found in literature is to use the effective wavelength $\lambda_{\mathrm{eff}}$ of the filter transmittivity curve to calculate the reddening correction. We demonstrate the implications in the following section.

\section{Implications}

\begin{figure}[t]
\includegraphics[width=6cm,height=3cm]{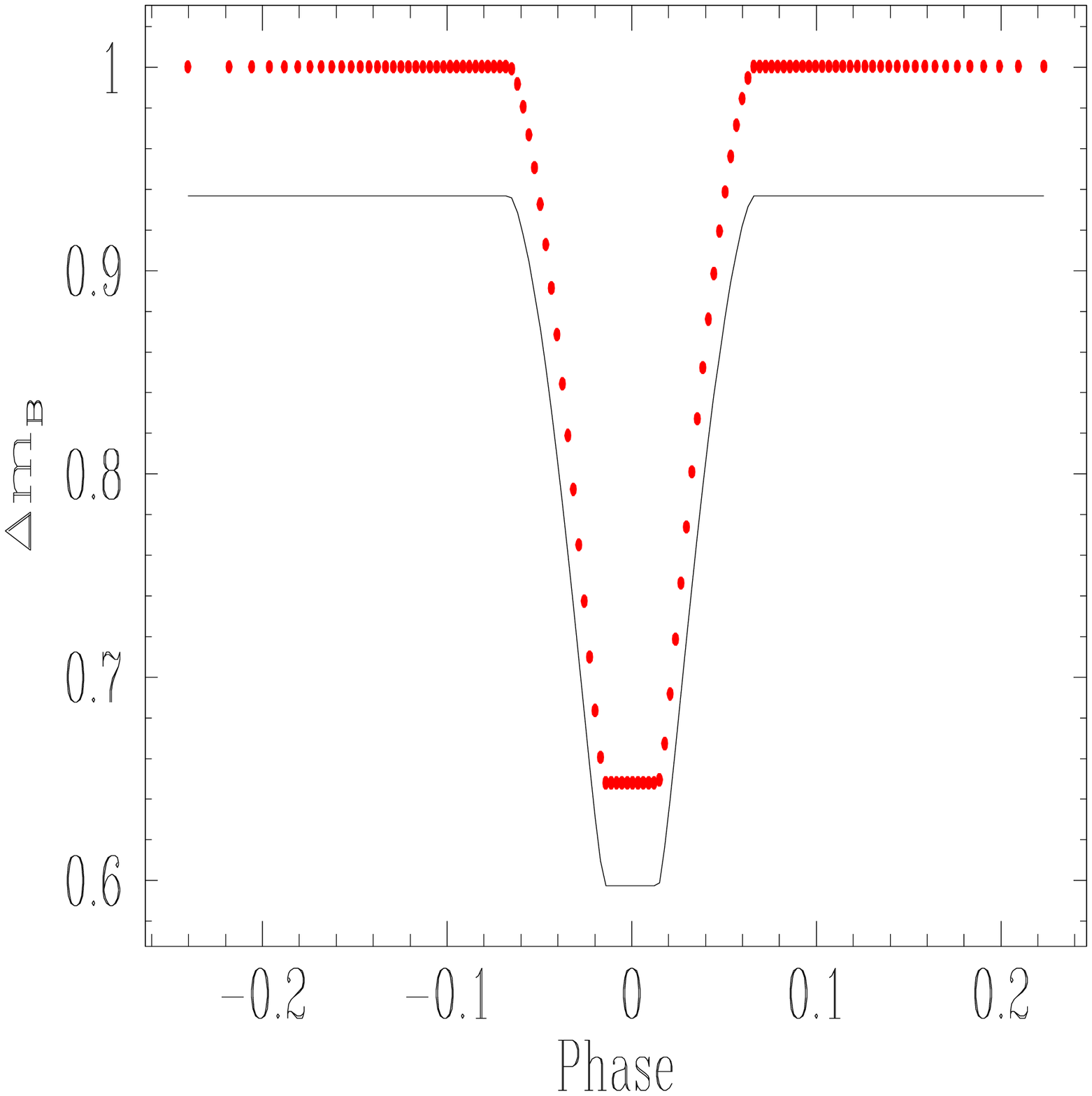}
\includegraphics[width=6cm,height=3cm]{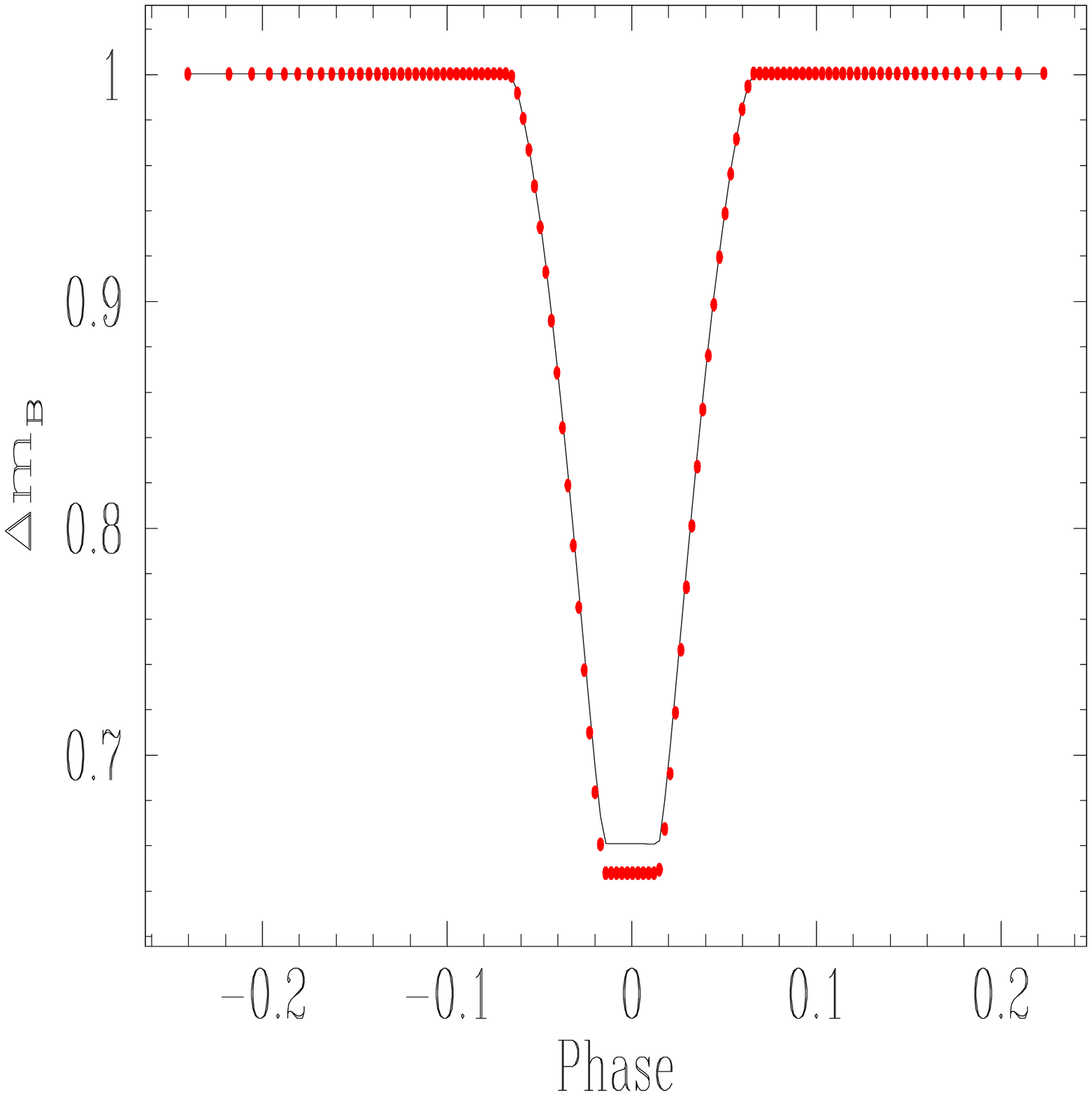} \\
\caption{Left: The discrepancy due to simplified constant subtraction approach (solid line) compared to the rigorously applied reddening (points) for a G9~V--F5~V binary. The subtracted constant was obtained from the effective wavelength ($\lambda_{\mathrm{eff}} = 4410.8$\AA) of the Johnson B transmittivity curve. Right: Overplotted light curves with the subtraction constant calculated so that the magnitudes in quarter phase are aligned. There is still a \emph{measurable} difference in eclipse depth of both light curves. $E(\mathrm B - \mathrm V) = 1$ is assumed.} \label{discrepancy}
\end{figure}

\begin{figure}[b]
\includegraphics[width=6cm,height=3cm]{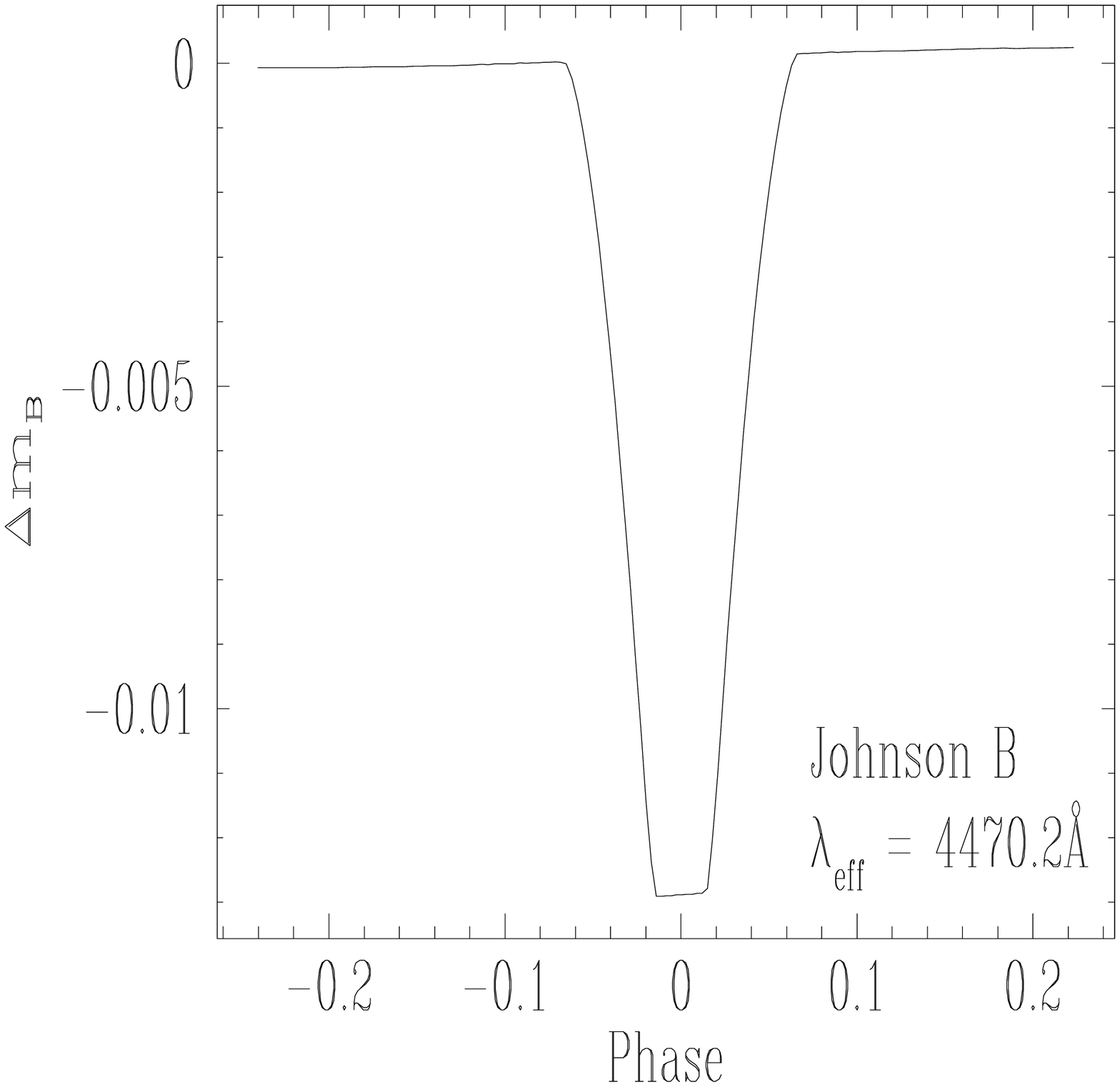}
\includegraphics[width=6cm,height=3cm]{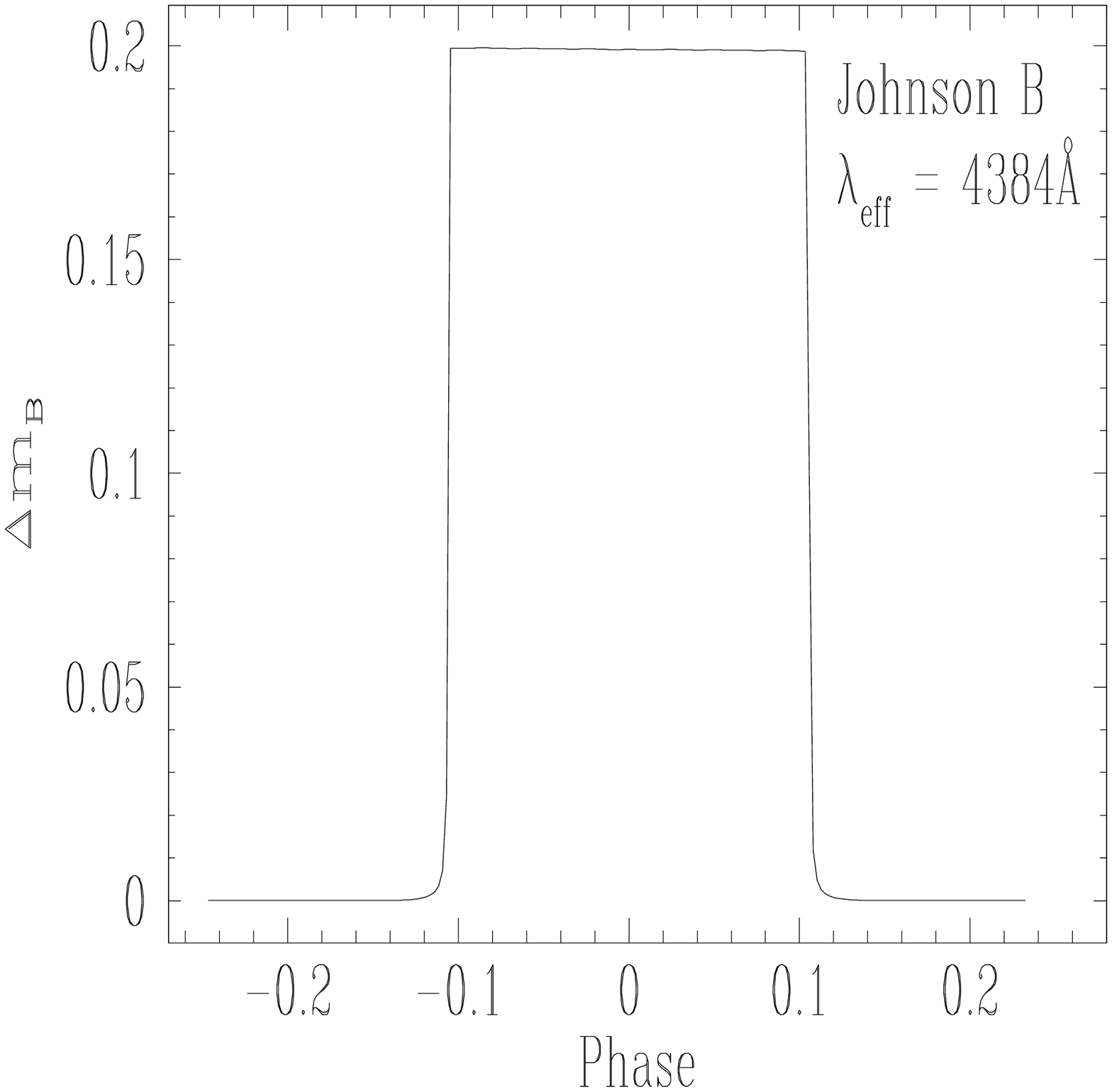} \\
\caption{The difference between the rigorously calculated reddening and a constant subtraction in Johnson B filter for the G9~V--F5~V binary (left) and a B8~V--K4~III binary (right) during primary minimum. $E(\mathrm B - \mathrm V) = 1$ is assumed.}
\label{minima}
\end{figure}

\subsection{Interstellar extinction}

By comparing the rigorously calculated fluxes against intrinsic fluxes with a simple constant subtracted, we come to the following conclusions: {\bf 1)} taking the effective wavelength of the filter bandpass should be avoided. Since the flux is the integral over the filter bandpass, $\lambda_{\mathrm{eff}}$ has a \emph{conceptually} different meaning. Furthermore, $\lambda_{\mathrm{eff}}$ of the given filter depends heavily on the effective temperature of the observed object and on the color excess $E(\mathrm B - \mathrm V)$ (Fig.~\ref{weff}). To determine the subtraction constant, one has to make sure that \emph {the integral} (rather than any particular wavelength) of the both curves is the same. Fig.~\ref{discrepancy} shows the discrepancy between the properly calculated light curve and the one obtained by subtracting a $\lambda_{\mathrm{eff}}$-calculated constant. Table \ref{analysis} summarizes the differences between the proper treatment and other approaches. {\bf 2)} Even if the subtraction constant is properly calculated, the light curves still exhibit measurable differences in both minima (Figs.~\ref{discrepancy}, \ref{minima}). This is due to the effective temperature change of the binary system during eclipses. For the analysed case, the difference in B magnitude is $\sim 0.01$mag, which is generally observable. {\bf 3)} If light curves in three or more photometric filters are available, it is possible to \emph{uniquely} determine the color excess value $E(\mathrm B - \mathrm V)$ by comparing different color indices in-and-out of eclipse. The reddening may thus be properly introduced to the fitting scheme of the eclipsing binary analysis program. This was done in {\tt PHOEBE}.

\begin{table}
\begin{tabular}{lccr}
Approach: & $\lambda_{\mathrm{eff}} $ & $\Delta m_B$ & $\varepsilon_{\Delta m_B}$ \\
\hline
Rigorously calculated value:              & 4470.2\AA & 4.03 &  0.00 \\
Filter transmittivity:                    & 4410.8\AA & 4.09 &  0.06 \\
Filter transmittivity + reddening law:    & 4452.1\AA & 4.05 &  0.02 \\
Intrinsic spectrum:                       & 4436.3\AA & 4.06 &  0.03 \\
Effective (reddened) spectrum:            & 4583.6\AA & 3.90 & -0.13 \\
\caption{The summary of different approaches to calculate the wavelength to be used for the dereddening constant. $\lambda_{\mathrm{eff}}$ is determined by requiring that the area under the spectrum on both sides is equal. $\Delta m_B$ is the value of extinction in B filter and $\varepsilon_{\Delta m_B}$ is the deviation from the rigorously calculated value. All values are calculated for $E(\mathrm B - \mathrm V) = 1$ at quarter phase. Note that $\Delta m_B$ is smaller than $\mathcal R_\mathrm V + E(\mathrm B - \mathrm V) = 4.1$, since our simulated binary is cooler than 10000K.}
\label{analysis}
\end{tabular}
\end{table}

\subsection{Atmospheric extinction}

\begin{figure}[ht]
\includegraphics[width=6cm,height=3cm]{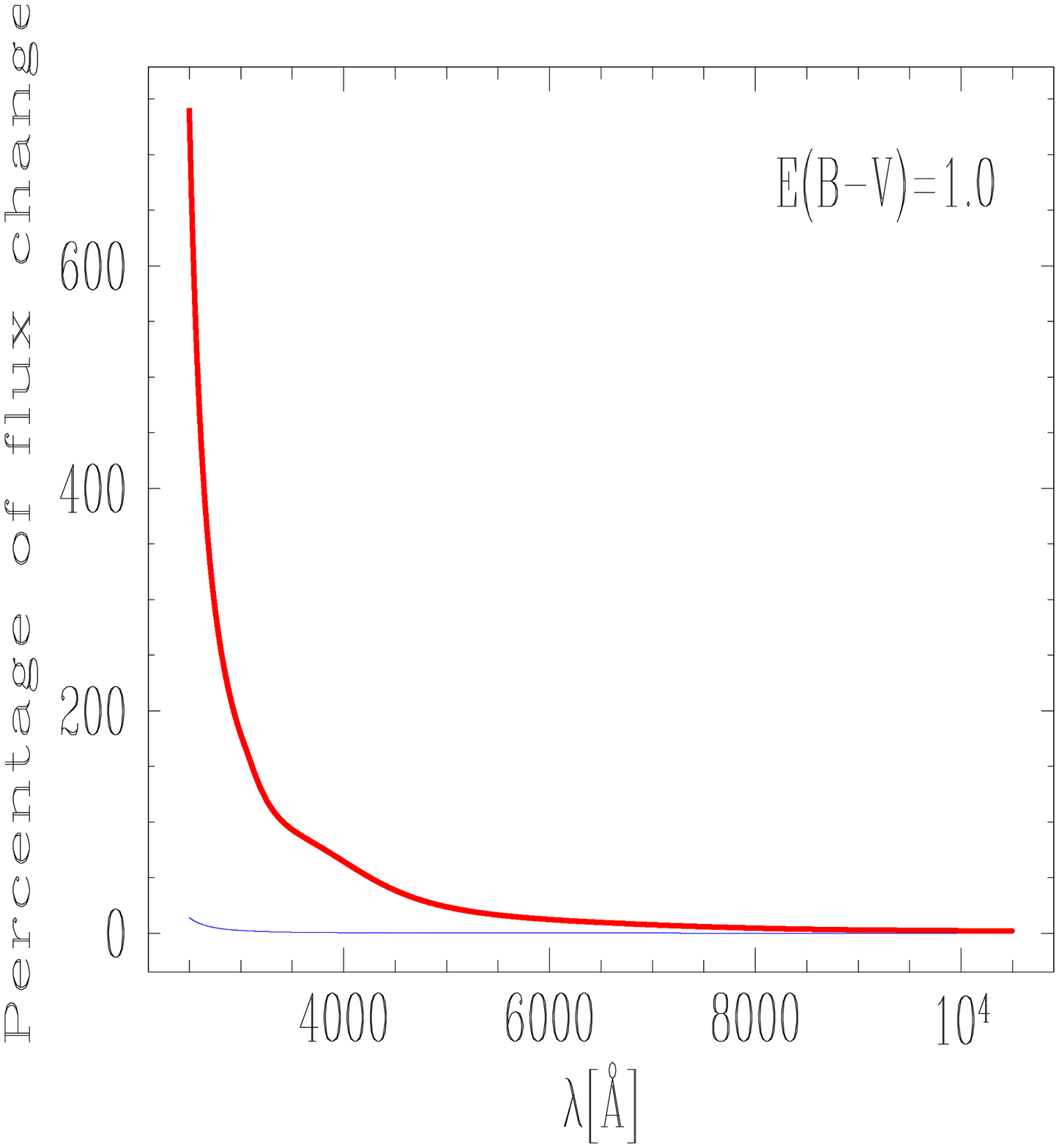}
\includegraphics[width=6cm,height=3cm]{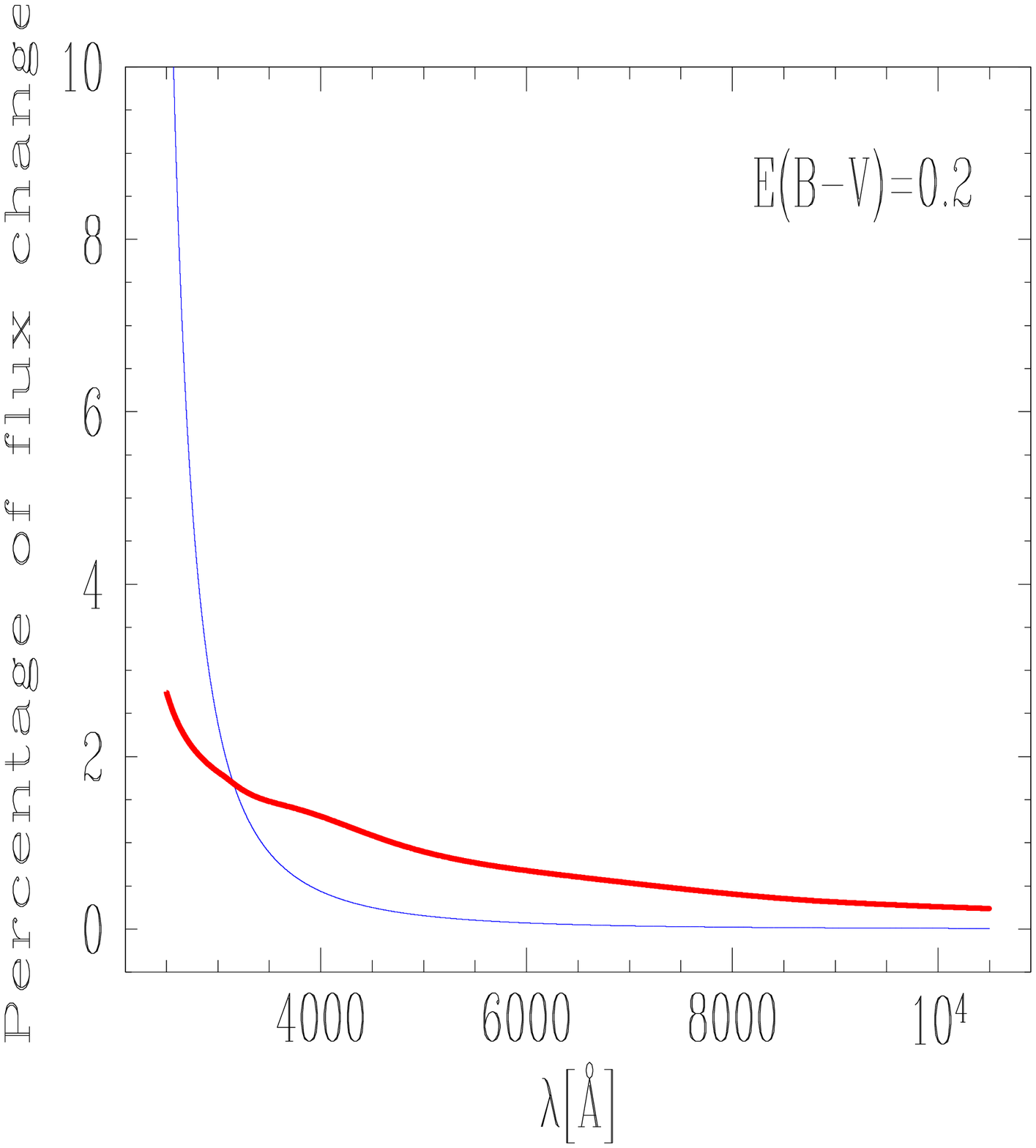} \\
\caption{The percentage of the flux change imposed on the intrinsic spectrum by reddening (thick line) and atmospheric extinction (thin line). If the color excess is large, the reddening completely dominates the spectrum (left), but if the color excess is moderate, the blue part of the spectrum is dominated by the Rayleigh scattering (right).}
\label{atmosphere}
\end{figure}

Atmospheric extinction is comprised of three different sources: the Rayleigh scattering, the aerosol scattering and ozone absorption \cite{forbes1996}. It depends on the wavelength of the observed light and on the air-mass of obsevations, which is of course the same for both binary components; we are thus left only with wavelength dependence at some inferred air-mass. To assess the impact on the photometric data, we compare the flux change imposed on the intrinsic spectrum by reddening and by atmospheric extinction (Fig.~\ref{atmosphere}). We conclude that for weakly and moderately reddened eclipsing binaries the atmospheric extinction dominates the blue parts of the spectrum due to Rayleigh scattering, but for larger color excesses ($E(\mathrm B - \mathrm V) \gtrsim 0.5$) the reddening is dominant throughout the spectrum.

\section{Discussion}

By not properly taking reddening into account, we introduce systematics of $\sim 0.01$mag into the solution for our simulated binary. One may ask: is this difference worth bothering with? To demonstrate that it is, let us simulate a bit more exotic eclipsing binary with blue B8 main-sequence ($T_1 = 12000$K, $R_1 = 3.221 R_\odot$) and red K4 type III giant ($T_2 = 4000$K, $R_2 = 30.0 R_\odot$) components with orbital period of 15 days. Fig.~\ref{minima} (right) shows that the discrepancy between the rigorously calculated light curve and the constant-subtracted one is as large as $\sim 0.2$mag, which means $\sim 10$\% error in the determined distance to the observed object. On the other hand, the reddening effect may be reduced by using narrower filter-sets, e.g.~Str\"omgren {\it ubvy} set. Furthermore, typical observed color excesses rarely exceed few tenths of the magnitude, which additionally diminishes this effect. However, future space scanning missions such as {\tt GAIA} \cite{perryman2001} will acquire measurements of eclipsing binaries without spatial bias. Thus interstellar extinction should be carefully and rigorously introduced into the reduction pipeline for eclipsing binaries.

\end{article}
\end{document}